\documentclass[aps,prb,twocolumn,superscriptaddress,groupedaddress,amsmath,amssymb,citeautoscript,floatfix]{revtex4}
\usepackage{graphicx}
\usepackage{hyperref}
\usepackage{nicefrac}
\usepackage[capitalize]{cleveref}
\usepackage{amsmath}
\usepackage{float}
\begin{document}
\title{Competition between intrinsic and extrinsic effects in the quenching of the superconducting state in FeSeTe thin films
}

\author{Antonio Leo}
\email{antoleo@sa.infn.it}
\affiliation{Dipartimento di Fisica ``E. R. Caianiello'', Universit\`a degli Studi di Salerno, I-84084 Fisciano (Salerno), Italy}
\affiliation{CNR-SPIN Salerno, I-84084 Fisciano (Salerno), Italy}

\author{Pasquale Marra} 
\email{pasquale.marra@spin.cnr.it}
\affiliation{CNR-SPIN Salerno, I-84084 Fisciano (Salerno), Italy}
\affiliation{Dipartimento di Fisica ``E. R. Caianiello'', Universit\`a degli Studi di Salerno, I-84084 Fisciano (Salerno), Italy}

\author{Gaia Grimaldi}
\affiliation{CNR-SPIN Salerno, I-84084 Fisciano (Salerno), Italy}
\affiliation{Dipartimento di Fisica ``E. R. Caianiello'', Universit\`a degli Studi di Salerno, I-84084 Fisciano (Salerno), Italy}

\author{Roberta Citro}
\affiliation{Dipartimento di Fisica ``E. R. Caianiello'', Universit\`a degli Studi di Salerno, I-84084 Fisciano (Salerno), Italy}
\affiliation{CNR-SPIN Salerno, I-84084 Fisciano (Salerno), Italy}

\author{Shrikant Kawale}
\author{Emilio Bellingeri}
\author{Carlo Ferdeghini}
\affiliation{CNR-SPIN Genova, Corso Perrone 24, I-16152 Genova, Italy}

\author{Sandro Pace}
\author{Angela Nigro}
\affiliation{Dipartimento di Fisica ``E. R. Caianiello'', Universit\`a degli Studi di Salerno, I-84084 Fisciano (Salerno), Italy}
\affiliation{CNR-SPIN Salerno, I-84084 Fisciano (Salerno), Italy}



\begin{abstract}
We report the first experimental observation of the quenching of the superconducting state in current-voltage characteristics of an iron-based superconductor, namely, in FeSeTe thin films.
Based on available theoretical models, our analysis suggests the presence of an intrinsic flux-flow electronic instability along with non-negligible extrinsic thermal effects.
The coexistence and competition of these two mechanisms classify the observed instability as halfway between those of low-temperature and of high-temperature superconductors, where thermal effects are respectively largely negligible or predominant.
\end{abstract}

\maketitle

\section{Introduction}

Flux-flow instabilities are abrupt transitions from the flux-flow regime of type-II superconductors to the normal state\cite{Larkin1975_Larkin1986}, observed in the current-voltage ($I$-$V$) characteristic at bias currents larger than the critical current $I_c$ and below the critical temperature $T_c$.
These instabilities have been extensively studied in low-temperature superconductors (LTSCs)\cite{Klein1985} and in cuprate high-temperature superconductors (HTSCs)\cite{Kalisky2006}.
Specifically, they can be used as a probe of quasiparticle energy-relaxation processes\cite{Xiao1999}, which are related to the pairing mechanism and superconducting gap\cite{Kaplan1976,Kalisky2006}, as well as to investigate the interplay between vortex dynamics and pinning\cite{Ruck1997}.
Moreover, as far as applications are concerned, their disruptive impact on the stability of the superconducting state is of primary importance to the realization and improvement of any superconducting device, e.g., magnets, current-leads, or fault-current limiters.

Whereas in LTSCs instabilities are generally driven by an intrinsic electronic mechanism\cite{Ruck1997,Peroz2005,Liang2010}, the superconducting quenching in HTSCs\cite{Kalisky2006,Doettinger1994} is more controversial in nature, due to the presence of non-negligible thermal effects\cite{Gurevich1987,Bezuglyj1992,Maza2008}, and observable only under very constrained experimental conditions, e.g., within narrow ranges of applied magnetic field and temperature.
Besides, the recently discovered iron-based superconductors\cite{Paglione2010,Chubukov2015} are very promising for technological applications, being major competitors of MgB$_2$ and cuprates\cite{Hosono2015Exploration}, due to their extraordinarily high upper critical fields\cite{Tanabe2012}, high current densities\cite{Matsumoto2014}, and high isotropies\cite{Her2015}.
From a fundamental point of view, these materials are unconventional superconductors which exhibit a distinctive multiband structure of the superconducting gap, and whose pairing mechanism is still under debate\cite{Chubukov2015}.
To date, there has been no experimental evidence of flux-flow instabilities in iron chalcogenides, to the best of our knowledge.
Very recently, a related phenomenon has been observed in a BaKFeAs mesoscopic structure, where the transition to the normal state due to non-magnetic impurities is ascribed to phase-slips\cite{Li2015}.

In this work, we report the first experimental observation of the quenching of the superconducting state associated to the flux-flow instability in the $I$-$V$ characteristics of FeSeTe thin films.
Although thermal effects have been minimized by a pulsed bias technique, their residual contributions have been taken into account by evaluating the temperature increase of the film due to Joule self-heating.
Nevertheless, experimental data in external magnetic field reveal the intrinsic nature of the instability, which therefore coexists with a non-negligible contribution of extrinsic thermal effects.

\begin{figure}
\includegraphics[width=\columnwidth]{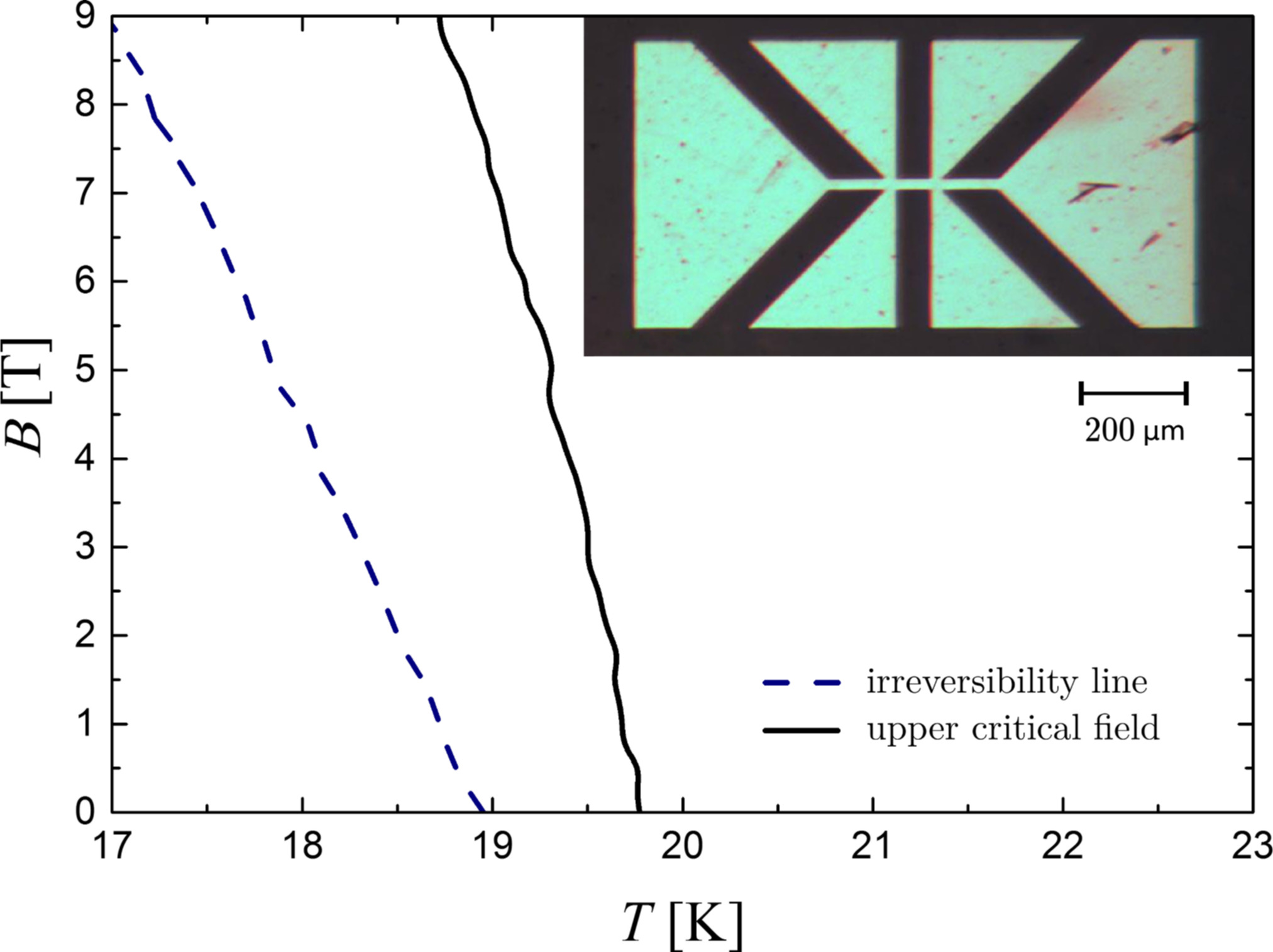}
\caption{
Irreversibility and upper critical field lines of the considered FeSeTe microbridges.
Inset: SEM image of the 20\,$\mu$m microbridge showing the typical geometry of the samples.
}
\label{fig:1}
\end{figure}

\section{The flux-flow instability}

In type-II superconductors, the quenching of the superconducting state occurs with a voltage jump in the $I$-$V$ characteristic at the instability current $I^*$, as a consequence either of an intrinsic mechanism related to the vortex dynamics\cite{Larkin1975_Larkin1986} or of extrinsic phenomena due to self-heating effects\cite{Vina2003_Xiao2001}.
The fingerprint of the intrinsic origin of the quenching is the presence of a distinctive shape of the $I$-$V$ characteristic in voltage-driven measurements.
In this case, a negative differential resistance appears at the instability current $I=I^*$ and voltage $V=V^*$, which subsequently merges the normal state branch $V=R_N I$, when the current reaches again the value $I\approx I^*$.
In current-driven measurements this feature is hidden by a sudden jump of the voltage at $I=I^*$.
Within the Larkin-Ovchinnikov (LO) model\cite{Larkin1975_Larkin1986}, flux-flow instabilities in $I$-$V$ curves at $V=V^*$ correspond to sudden drops in the viscosity of the Abrikosov lattice due to the shrinking of the vortex cores, which occurs at the critical velocity $v^*=V^*/(BL)$, with $B$ being the applied magnetic field and $L$ the distance between voltage contacts.
The LO model describes the instability both in LTSCs\cite{Peroz2005,Leo2011} and HTSCs\cite{Stoll2000}, whenever the basic assumptions of the model are satisfied, i.e., temperature near $T_c$ and small magnetic field, in which the quasiparticle distribution can be assumed spatially uniform.
In its original formulation, the LO model predicts an almost field-independent critical vortex velocity $v^*$.
The experimental evidence of a field-dependent critical velocity can be described by an empirical extension of the original model\cite{Doettinger1994}.
At low applied magnetic fields, it has been found that $v^*\propto B^{-1/2}$, as a direct consequence of the spatial homogeneity of the non-equilibrium quasiparticle distribution in the surrounding of Abrikosov vortices\cite{Doettinger1994}.
Moreover, the LO model predicts a monotonic decrease of temperature-dependent vortex velocity $v^*$ near $T_c$.
However, substantial deviations can be observed due to the competition between quasiparticle scattering and recombination processes\cite{Peroz2005,Leo2011}.

At temperatures well below $T_c$ instead, the electronic instability can be related to the electron overheating\cite{Kunchur2002,Knight2006}.
Due to dissipations in the flux-flow regime, electrons reach temperatures above the phonon temperature, which leads to an expansion of the vortex core and a consequent reduction of the viscous drag.
The signature of the instability in the $I$-$V$ curves is very similar in the LO and in the electron overheating regime, despite the different mechanisms present in the two distinct temperature ranges.
Recently, a crossover from electron heating at low temperatures to the LO regime near $T_c$ has been observed in nanostructured $a$-NbGe films\cite{Otto2010}, at the reduced temperature $T/T_c=0.68$.
However, the magnetic-field dependence of the critical vortex velocity is $v^*\propto B^{-1/2}$ in the two temperature regimes\cite{Kunchur2002}.
Furthermore, regardless of the microscopic mechanism, a more complex dependence can be present at very low fields, due to the pinning properties of the material and sample geometry\cite{Grimaldi2010_Silhanek2012_Grimaldi2012_Grimaldi2015}.

\begin{figure}
\includegraphics[width=\columnwidth]{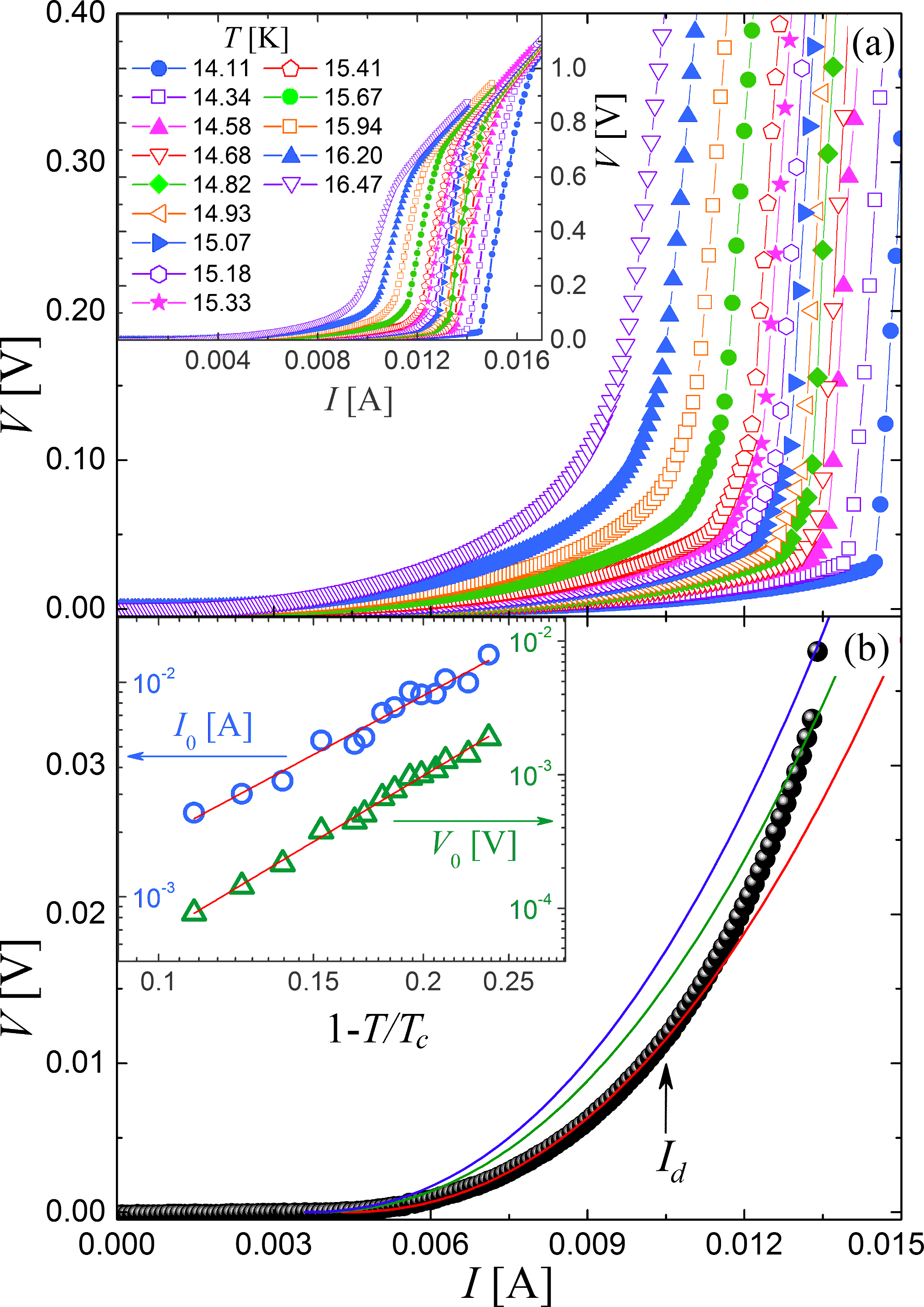}
\caption{
(a) $I$-$V$ characteristics of FeSeTe microbridge at different temperatures in a magnetic field $B=5$\,T.
Inset: $I$-$V$'s in the full voltage range.
Solid lines are guides for the eye.
(b) Experimental $I$-$V$ characteristic at 14.58\,K.
Solid lines are isothermal curves from \cref{eq:VI}, with $T=$14.58\,K, 14.82\,K, and 14.93\,K.
Inset: Temperature dependence of the critical current $I_c$ (left scale) and of the fitting parameter $V_0$ (right scale) as a function of $1-T/T_c$.
Solid lines are best fits of \cref{eq:V00Ic0}.
}
\label{fig:2}
\end{figure}

The quenching of the superconducting state can also occur due to the Joule self-heating caused by thermal dissipations of the vortex flow.
The occurrence of this thermal instability in HTSCs at zero applied magnetic field has been theoretically described as a pure consequence of a temperature runaway which takes place up to the quenching temperature $T^*$, where a very steep voltage jump sets in\cite{Maza2008,Maza2011}.
It is claimed that, under certain experimental conditions, heating effects become predominant\cite{Maza2008}, and at sufficiently high current densities, cuprate HTSCs become thermally unstable\cite{Maza2011}.
On the other hand, the LO model has been extended by Bezuglyj and Shklovskij\cite{Bezuglyj1992} (BS) in order to include the effect of self-heating near $T_c$ in a finite magnetic field.
In this case instead, a crossover scenario emerges, with a threshold value of the applied magnetic field $B_T$ between the LO and pure self-heating regimes.
For $B\ll B_T$ the instability is triggered by an intrinsic change of the distribution function of quasiparticles trapped in the vortex core, whereas for $B\gg B_T$ it is driven by pure thermal effects\cite{Bezuglyj1992}.
This crossover corresponds to a distinctive magnetic-field dependence of the Joule power $P^*=I^*V^*$ at the instability, which is linear at low fields $B\ll B_T$ and saturates for $B\gg B_T$.

In general, the relative importance of thermal effects can be characterized by the value of the Stekly parameter\cite{Gurevich1987} $\alpha$, defined as the ratio of the heat generation in the normal state $\rho_N J_c^2$ and the heat transfer $2(T_c-T)h/D$, being $\rho_N$, $J_c$, $h$, and $D$ respectively the normal state resistivity, critical current density, heat transfer coefficient, and film thickness\cite{Gurevich1987}.
Self-heating is important if $\alpha>1$, and negligible if $\alpha\ll1$.
This parameter gives a measure of thermal contributions to the flux-flow instability, and depends on the intrinsic properties of the sample, but also on experimental conditions and on the film substrate.
In light of the above theoretical discussion, we will try to characterize the nature of the instability in the iron-based superconductor FeSeTe through a detailed analysis of the experimental $I$-$V$ characteristics.

\section{Experiment}

The FeSeTe thin films have been grown on CaF$_2$ substrates by pulsed laser deposition from a FeSe$_{0.5}$Te$_{0.5}$ target\cite{Bellingeri2011}.
The resulting film thickness is $D=120$\,nm.
A four-probe pattern has been realized in form of microbridges with length $L=65$\,$\mu$m and width $W=20$\,$\mu$m, by standard UV photolithography and Ar ion-milling etching.
The inset of \cref{fig:1} shows the typical geometry of our samples.
From the superconducting phase diagram reported in \cref{fig:1}, it results that the typical $T_c$ of the samples is about 20.5\,K, with a transition width $\Delta T_c\le1$\,K.
The $T_c$ has been evaluated as the temperature value corresponding to 50\% of the normal state resistance at zero applied magnetic field.
The transition width is defined as the difference between the temperature values corresponding to 90\% and 10\% of the normal state resistance.
These two criteria define respectively the upper critical field $H_{c2}$ and the irreversibility line of the phase diagram, shown in \cref{fig:1}.
Furthermore, the critical current density is $J_c\approx5\times10^5\,{\rm A}/{\rm cm}^2$ at 15\,K.
More information about sample fabrication and their structural and pinning properties can be found in Refs.~[\onlinecite{Braccini2013_Bellingeri2014_Leo2015}].
The $I$-$V$ characteristics have been acquired in a cryogen-free cryostat equipped with an integrated cryogen-free variable-temperature insert operating in the range 1.6--300\,K and a superconducting magnet able to generate fields up to 9\,T.
The sample is thermally coupled to a copper block in flowing He gas, with a temperature stability within 0.01\,K.
In order to minimize self-heating effects, $I$-$V$ curves have been acquired by a pulsed current technique, with rectangular pulses of duration 2.5\,ms.
The inter-pulses current-off time has been set to 1\,s to ensure the complete re-thermalization to the He flow temperature $T$.
The magnetic field is applied perpendicularly to the film surface.

\section{Results and discussion}

\begin{figure}
\includegraphics[width=\columnwidth]{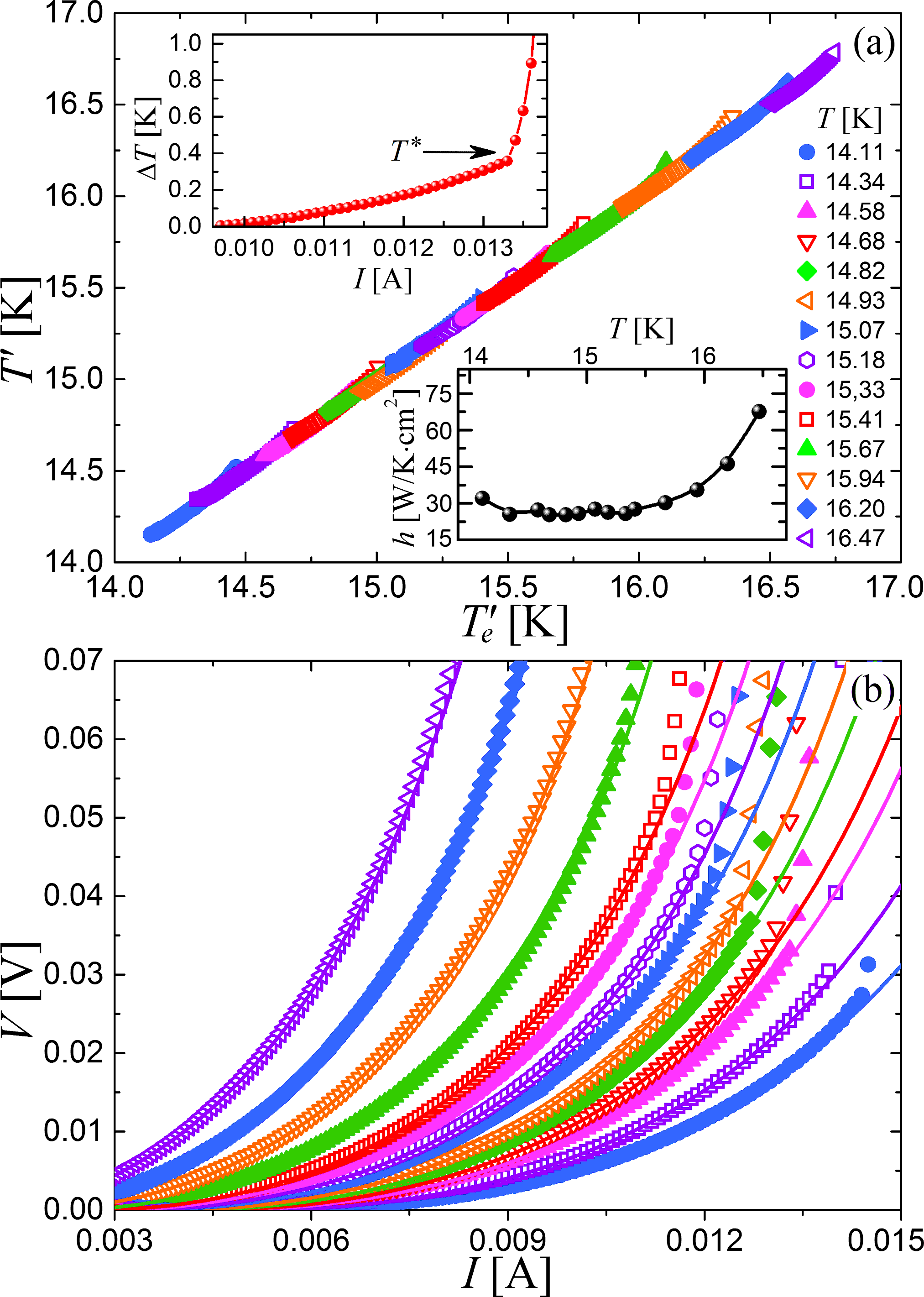}
\caption{
(a) Sample temperature $T'$ as a function of its estimate $T'_e$ for all experimental $I$-$V$ curves.
Top inset: $\Delta T$ as a function of the bias current $I$.
Bottom inset: Temperature dependence of the heat-transfer coefficient $h(T)$.
(b) Non-isothermal $\widetilde{V}(I,T)=V(I,T'_e)$ (solid lines) fitted to the experimental $I$-$V$'s (symbols).
}
\label{fig:3}
\end{figure}

\Cref{fig:2} shows the measured $I$-$V$ curves in the temperature range 14.0--16.5\,K with an applied magnetic field of 5\,T.
No hysteresis has been observed by sweeping the bias current up to the maximum value and then down to 0\,A.
A voltage jump from the flux-flow regime to the normal state occurs at temperatures below 15.5\,K.
Note, however, that the observed jump is not abrupt, but smoothed out by the presence of extrinsic thermal effects, as we will show below.
We also remark that several other samples with larger microbridge width ($W\gg20\,\mu$m) have been measured, but no jump has been observed, probably due to a larger self-heating.
In the low-current regime, where self-heating processes are negligible, the measured temperature of the bath $T$ is a good estimation of the actual temperature of the sample.
Within this assumption, it is possible to employ the functional form suggested in Ref.~[\onlinecite{Maza2008}] to fit experimental data at low bias above the critical current $I_c$.
In this regime, the measured $I$-$V$ characteristics are indeed well described by the isothermal curves
\begin{gather}
V(I,T) = V_0(T) \left[ \frac{I}{I_c(T)}-1 \right]^n,
\quad\text{with}\quad
\label{eq:VI}
\\
I_c(T) = I_{c0} 
\left( 1-\frac{T}{T_c} \right)^p, \quad
V_0(T) = V_{00} 
\left( 1-\frac{T}{T_c} \right)^q,
\label{eq:V00Ic0}
\end{gather}
where $T_c=18.5$\,K at $B=5$\,T.
As an example, \cref{fig:2}(b) shows the fitting curve from \cref{eq:VI} to the experimental $I$-$V$ at 14.58\,K.
The values of $V_0$, $I_c$, and $n$ have been fitted to the experimental data at low bias.
At low bias, \cref{eq:VI} reproduces the experimental data with constant exponents $n$, $p$, and $q$ in the full temperature range 14--16.5\,K.
This can be revealed by the temperature dependence of $V_0$ and of the critical current $I_c$, shown in the inset of \cref{fig:2}(b).
In particular, we found $n=2$, $p=2.5$, $q=3.1$, $I_{c0}=0.19$\,A, and $V_{00}=0.64$\,V.

\begin{figure}
\includegraphics[width=\columnwidth]{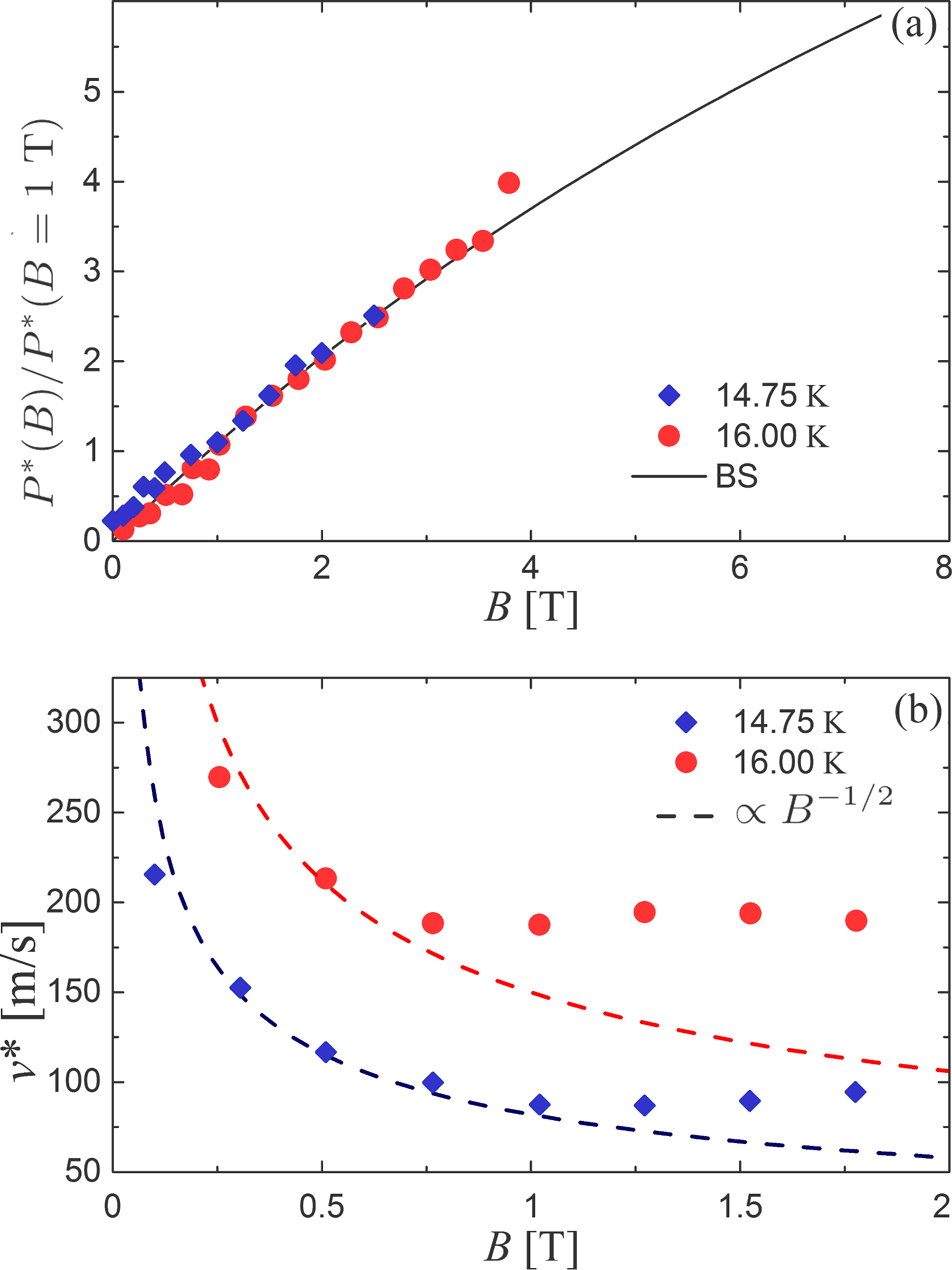}
\caption{
(a) Normalized Joule power $P^*(B)/P^*(B=1\text{\,T})$ at temperatures $T=$14.75\,K and 16\,K.
The solid line is the predicted Bezuglyj-Shklovskij\cite{Bezuglyj1992} (BS) dependence with $B_T=25$\,T.
(b) Critical vortex velocity $v^*$ as a function of applied magnetic field $B$ for both temperatures.
Solid lines are best fits of $v^*\propto B^{-1/2}$ at low fields.
}
\label{fig:4}
\end{figure}

As the bias current increases, the experimental $I$-$V$ curve departs from the isothermal curve at a threshold value $I_d$, as shown in \cref{fig:2}(b).
This discrepancy can be explained as the effect of the self-heating\cite{Maza2008,Maza2011} for currents $I > I_d$.
In this current regime, the actual temperature of the sample $T'$ does not correspond anymore to the measured temperature $T$.
One can estimate the temperature $T'$ by fitting experimental data points $(V,I)$ with the isothermal curves $V(I,T)$ in \cref{eq:VI}.
The temperature difference $\Delta T=T'-T$ increases smoothly as the current increases up to the quenching temperature $T^*$, where $\Delta T$ shows a jump, as shown in the top inset of \cref{fig:3}.
This behavior is consistent with the analysis performed in Ref.~[\onlinecite{Maza2008}] for HTSC thin films (YBCO).
However, in our case, the temperature increase $\Delta T$ is smaller than $0.4$\,K below the instability.
This difference can be ascribed to the different thermal properties of YBCO and FeSeTe, but also to the difference between our pulsed biasing technique and the ramp method used in Ref.~[\onlinecite{Maza2008}].
For the specific pulsed technique employed, the temperature increase $\Delta T$ can be easily related to power dissipations in the flux-flow regime.
In fact, during a single rectangular pulse the current can be assumed as uniformly distributed over the sample cross section.
Moreover, due to the long current-off time used between two consecutive current pulses, the initial temperature of the sample can be assumed to coincide with the bath temperature $T$ for all pulses.
Under these assumptions, the steady-state heat balance yields $T'=T+(VI/C)$, i.e., the temperature increase $\Delta T$ is proportional to the Joule power dissipation $P=VI$.
The parameter $C$ is proportional to the heat-transfer coefficient $h$ between the sample and the cooling environment\cite{Gurevich1987}, i.e., $C=2L(W+D)h$.
The Joule power $P=VI$ can be directly calculated from the isothermal $I$-$V$ characteristics, and compared with the temperature increase $\Delta T$.
We find indeed that the temperature $T'$ is a linear function of the power $P$.
This can be recognized in \cref{fig:3}(a), where $T'$ is compared with its estimate $T'_e$ obtained as the linear fit of $T'$ versus $P$, for all experimental $I$-$V$ curves.
Indeed, all data points line up along a line of unitary slope.
Hence, one can obtain the theoretical non-isothermal characteristic as $\widetilde{V}(I,T)=V(I,T'_e)$.
This theoretical $I$-$V$ characteristic takes explicitly into account the effect of the self-heating and can be eventually compared with the experimental $I$-$V$ curves, as shown in \cref{fig:3}(b).
Solid lines are best fits of the theoretical non-isothermal $\widetilde{V}(I,T)$ dependence to the experimental curves, below the instability current $I^*$.
The parameter $C(T)$, inferred by the best fit of the temperature $T'$ as a function of the power $P(T)$, yields the temperature dependence of the heat-transfer coefficient $h(T)$ reported in the bottom inset of \cref{fig:3}(a).
A quite constant value $h\approx 30$\,W/cm$^2\cdot$K is observed up to 16\,K, which gives a Stekly parameter $\alpha\approx 0.4$.
This value corresponds to an intermediate regime where self-heating is neither negligible ($\alpha\ll1$) nor overcomes intrinsic effects ($\alpha>1$)\cite{Gurevich1987}.
We remind that the value of the Stekly parameter strongly depends on material properties, film thickness, and on the heat flow at the film/bath and film/substrate interfaces, which is described by the heat transfer coefficient $h$.
In particular the heat flow at the film/substrate interface is mainly determined by its thermal resistance, which is affected by the quality of the interface, i.e., disorder and roughness, and by the differences in electronic and phononic properties between the film and substrate.
Different cooling environments and substrates, as well as the current bias technique and the quality of the sample, may change the value of the Stekly parameter and hence the relative importance of the thermal effects on the voltage instability in superconducting microbridges.
In the case of competing intrinsic and thermal effects, the analysis of experimental data at different applied magnetic fields can give valuable information.
\Cref{fig:4}(a) shows the power $P^*$ at the instability as a function of applied magnetic field at temperatures $T=$14.7\,K and 16\,K.
According to the BS model\cite{Bezuglyj1992}, the power $P^*$ increases linearly as the field increases, saturating to a constant value for $B\gg B_T$.
Our experimental data show indeed a linear increase at low fields, consistent with a value of $B_T> 25$\,T.
The investigated magnetic field range is thus well below the threshold field $B_T$, at which pure thermal mechanisms start to dominate.

The intrinsic nature of the instability is moreover consistent with the magnetic-field dependence of the vortex velocity $v^*=V^*/(BL)$.
As reported before, an intrinsic instability induces a characteristic field-dependence of the critical vortex velocity $v^*\propto B^{-1/2}$, in a wide range of temperatures $T<T_c$.
\Cref{fig:4}(b) shows the critical vortex velocity as a function of magnetic field at temperatures $T=$14.7\,K and 16\,K.
At low fields, the critical vortex velocity decreases with increasing field with the characteristic dependence $v^*\propto B^{-1/2}$, which is consistent with the intrinsic nature of the instability.
At larger fields, the velocity remains approximately constant in the range investigated.
We notice that the measured vortex velocities are at least one order of magnitude lower than those observed in LTSCs\cite{Peroz2005}, and comparable with those of HTSCs\cite{Kalisky2006}.

\begin{figure}
\includegraphics[width=\columnwidth]{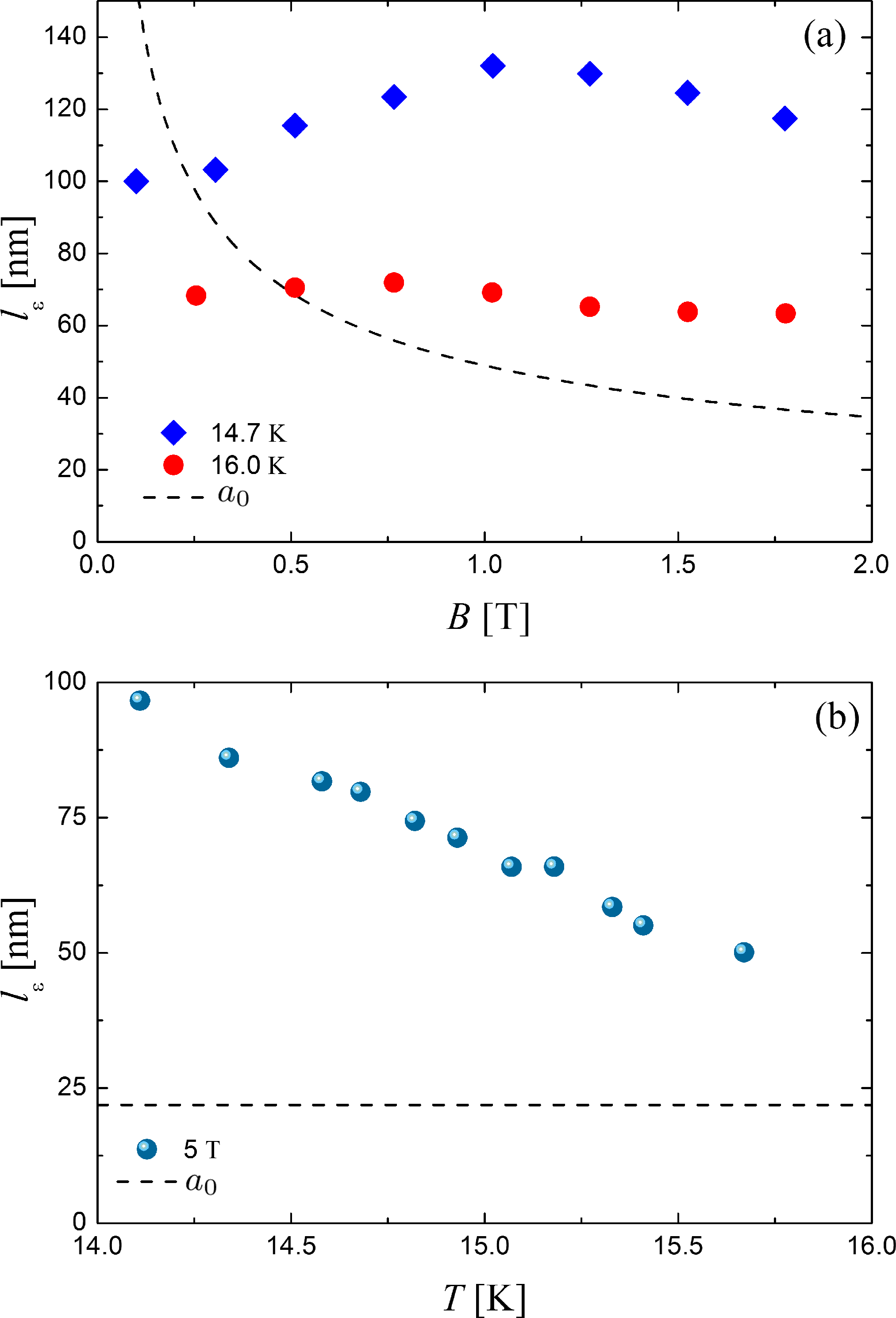}
\caption{
(a) The diffusion length of quasiparticles $l_{\varepsilon}$ as a function of applied magnetic field $B$ at temperatures $T=$14.75\,K and 16\,K.
The dashed line is the intervortex distance $a_0$ as a function of $B$.
(b) The diffusion length $l_{\varepsilon}$ as a function of temperature at $B=5$\,T.
The dashed line is the intervortex distance at $B=5$\,T.
} 
\label{fig:5}
\end{figure}

Moreover, other important physical parameters can be inferred from the analysis of flux-flow instability, in particular the quasiparticle scattering time $\tau$ and the diffusion length of quasiparticles $l_{\varepsilon}$\cite{Xiao1998}.
The diffusion length $l_{\varepsilon}$ is evaluated as\cite{Doettinger1995} $l_{\varepsilon} =(l_{\varepsilon0}/2)(1+\sqrt{1+4a_0/l_{\varepsilon0}})$, where $l_{\varepsilon0}=D_{qp} {\left[ 14\zeta(3)(1-T/T_c) \right]}^{1/4}/(\sqrt{\pi} v^*)$, $\zeta(x)$ is the Riemann function, $D_{qp}$ is the quasiparticle diffusion coefficient, and $a_0$ the intervortex distance.
The diffusion coefficient $D_{qp}$ is obtained from the slope ${\left[{\rm d}H_{c2}/{\rm d}T\right]}\vert_{T=T_c}=-10\,{\rm T}/{\rm K}$ of the upper critical field near $T_c$ as\cite{Tinkham} $D_{qp}=[4k_B/(\pi e)]/{\left[{\rm d}H_{c2}/{\rm d}T\right]}\vert_{T=T_c}$.
The intervortex distance is instead given by $a_0={\left[2\Phi_0/(\sqrt{3}B)\right]}^{{1}/{2}}$, where $\Phi_0$ is the magnetic flux quantum.
\Cref{fig:5} shows a comparison of $l_{\varepsilon}$ with the intervortex distance $a_0$.
We obtain that $l_{\varepsilon}>a_0$ within the full temperature range investigated and at magnetic fields above 0.5~T.
This value assures the homogeneity of the quasiparticle distribution over the whole volume of the superconductor\cite{Xiao1998}, which is a necessary condition for the applicability of the LO theory\cite{Larkin1975_Larkin1986}.
The quasiparticle scattering time can be obtained from its relation with the diffusion length\cite{Doettinger1995} $l_{\varepsilon}=\sqrt{D_{qp}\tau}$, which gives $\tau\approx10^{-9}$\,s.
This value is comparable with that obtained for YBCO\cite{Kunchur2002}.

\section{Conclusions}

We reported the observation of the quenching from the flux-flow regime to the normal state in $I$-$V$ characteristics of FeSeTe microbridges.
We find evidence of an intrinsic electronic instability which coexists with a non-negligible contribution of extrinsic thermal effects.
In particular, our theoretical analysis of the Joule self-heating suggests a temperature increase $\Delta T<0.4$\,K, whose current-dependence is consistent with a steady-state heat balance regime.
We thus infer a value of the Stekly parameter $\alpha\approx0.4$, which classifies FeSeTe as halfway between LTSCs and HTSCs, where thermal effects are respectively largely negligible or predominant.
Moreover, the value obtained for the quasiparticle scattering time $\tau$ is comparable with that of YBCO\@.
On top of that, the observed magnetic-field dependence of the critical vortex velocity and of the corresponding Joule power are consistent with an intrinsic nature of the instability.

\begin{acknowledgments}
A.\,L. acknowledges financial support from PON Ricerca e Competitivit\`a 2007--2013 under Grant Agreement PON NAFASSY, PONa3\_00007.
P.\,M. and R.\,C. acknowledge the project FIRB-2012-HybridNanoDev (Grant No. RBFR1236VV).
\end{acknowledgments}

\renewcommand{\mathrm}{}

\end{document}